\documentstyle[12pt]{article}
\begin{document}
\title{Contravariant symbol quantization on $S^2$}
\author{A.V.Karabegov\\
Joint Institute for Nuclear research, Dubna, Russia}
\maketitle
\begin{abstract}
We define an algebra of contravariant symbols on $S^2$ and give
an algebraic proof of the Correspondence Principle for that
algebra.
\end{abstract}
{\bf \S 0. Introduction.}

\bigskip

In \cite{GCQ} F.A.Berezin introduced a general
concept of quantization on a symplectic manifold $\Omega$.
To define a quantization on $\Omega$ one needs the following data.
Let $F$ be a set of positive numbers with a limit point 0.
For each $h\in F$ let ${\cal A}_h\subset C^\infty(\Omega)$
be an algebra with multiplication $\ast_h$ such that for
$h<h',\ {\cal A}_h\supset{\cal A}_{h'}$ as linear spaces. Denote
${\cal A}=\cup{\cal A}_h$. Assume that for each $h\in F$ there is
given a representation of ${\cal A}_h$ in some Hilbert space $H_h$.
These data define a quantization on $\Omega$
if the Correspondence Principle holds, i.e. for $f,g\in{\cal A}$
$$
 \lim_{h\to 0}f\ast_hg=fg;
 \quad \lim_{h\to 0}h^{-1}(f\ast_hg-g\ast_hf)=i\{f,g\},
$$
where $\{\cdot,\cdot\}$ denotes a Poisson bracket on $\Omega$.
If for $f\in {\cal A}_h\quad \hat f$ denotes a corresponding
operator in $H_h$, the function $f$ is called a symbol of $\hat f$.

Thus to define a quantization one may start from an appropriate construction
of symbols.  In \cite{CCS} Berezin introduced covariant and contravariant
operator symbols and extensively investigated their various properties.
In \cite{QKM} he applied covariant symbols to quantization of K\"ahler
manifolds.  A particular example of covariant symbol quantization on $S^2$
was considered in \cite{GCQ}. Therein Berezin described the algebra of
covariant
symbols and gave an analytic proof of the Correspondence Principle
for covariant symbol quantization on $S^2$.

A more advanced approach to quantization as to deformation of classical
mechanics was developed in \cite{B}.

In this paper we will define algebras of co- and contravariant symbols
on $S^2$, two of them in the same framework and give an algebraic proof of the
Correspondence Principle both for co- and contravariant symbol quantizations.

\bigskip

{\bf \S 1. Covariant and contravariant symbols on $S^2$.}

\bigskip

Consider a Hilbert space $L^2({\bf C}, d\alpha_n)$, with a measure
$$
    d\alpha_n(z,\bar z)=
\frac{n}{2\pi i}\frac{dz\wedge d\bar z}{(1+z\bar z)^{n+1}}
$$
and with the scalar product denoted by $(\cdot,\cdot)$.  Let $H_n$
be the n-dimensional subspace of $L^2({\bf C}, d\alpha_n)$ of all
polynomials in $z$ of degree  $\leq n-1$. For $v\in{\bf
C}$ the vectors $e_{\bar v}(z)=(1+z\bar v)^{n-1}\in H_n$ have
a following reproducing property, for
$f\in H_n\qquad f(v)=(f,e_{\bar v})$.

Definition.{\it The covariant symbol of an operator $A\in H_n$
is the function }
$f(z,\bar z)=(Ae_{\bar z},e_{\bar z})/(e_{\bar z},e_{\bar z})$.

To define a contravariant symbol one needs a notion of the canonical
measure $\mu_n$ on ${\bf C}$,
$$
    d\mu_n(z,\bar z)=(e_{\bar z},e_{\bar z})d\alpha_n(z,\bar z)=
\frac{n}{2\pi i}\frac{dz\wedge d\bar z}{(1+z\bar z)^2}.
$$
Let $P_{z,\bar z}$ denote the orthogonal projection operator on
$e_{\bar z}$ in $H_n$.

Definition.{\it A function $f(z,\bar z)$ is a contravariant symbol
of the operator $A$ if}
$$
           A=\int f(z,\bar z)P_{z,\bar z}d\mu_n(z,\bar z).
$$

Let $G=SU(2)$ be the group of all unitary $2\times 2$-matrices
with the determinant 1. Let $G$ act on ${\bf C}$ from the right by
fractional-linear transformations, for

\begin{equation}
g=\left(\begin{array}{cc}
a&b\\-\bar b&\bar a\end{array}\right)
\in G ,  \label{a}
\end{equation}

$z\in {\bf C}$
$g:z\mapsto zg=(az-\bar b)/(bz+\bar a)$. Actually $G$ acts on the
widened complex plane, i.e. on the Riemann sphere $S^2$ by rigid
rotations.  The canonical measure $\mu_n$, considered as a measure
on $S^2$, is rotation-invariant.

  For each natural $n$
 the group $G$ has exactly one unitary n-dimensional representation up
 to unitary equivalence.  Denote it by $\pi_n$. There exists a
 realization of $\pi_n$ in $H_n$ as follows. For
 $g$ given by (1)
and $f\in H_n$, one has
 $(\pi_n(g)f)(z)=(bz+\bar a)^{n-1}f(zg)$.
Since
 $\pi_n(g)e_{\bar v}=(\bar a-\bar b\bar v)^{n-1}e_{\overline{vg}^{-1}}$,
one immediately finds that both for covariant and contravariant symbols,
the symbol --- operator correspondence is $G$-equivariant, i.e. if
$f(z,\bar z)$ is a symbol of an operator $A$ in $H_n$ then
$f(zg,\overline{zg})$ is a symbol of $\pi_n(g)A\pi_n(g^{-1})$.
Thus it is natural to consider both covariant and contravariant symbols
as functions on $S^2$. In particular to define a covariant symbol
at infinity one needs to replace $e_{\bar z}$  by $e_\infty=z^{n-1}$
in the definition of a covariant symbol.
A nice invariant way to introduce the so called coherent states
$\lbrace e_{\bar v}\rbrace$
and covariant symbols in terms of line bundles can be found in \cite{RCG}.

\bigskip

{\bf \S 2. Symbols corresponding to the universal enveloping
algebra elements.}

\bigskip

The Lie algebra $su(2)$ of $G$ consists of all skew Hermitian
traceless $2\!\times\! 2$-matrices. Its complexification is the Lie algebra
$sl(2,{\bf C})$ of traceless complex $2\!\times\! 2$-matrices. Let $U$
denote the universal enveloping algebra of $sl(2,{\bf C})$.
The representation $\pi_n$ can be defined on $su(2)$ by derivation,
then extended to $sl(2,{\bf C})$ by complex-linearity and finally
extended to $U$. For
\begin{displaymath}
X=\left(\begin{array}{cc}
a&b\\c&-a\end{array}\right)
\in sl(2,{\bf C})
\end{displaymath}
one has
\begin{equation}
\pi_n(X)=(-bz^2+2az+c)\frac{d}{dz}+(n-1)(bz-a).
\label{b}
\end{equation}
Thus for $u\in U$ $u_n=\pi_n(u)$ is a differential operator
with polynomial coefficients in $z$ and $n$. Let $s_n(u)$ denote
the covariant symbol of $u\in U$. It can be calculated as follows
\begin{equation}
s_n(u)=\frac{(u_ne_{\bar z},e_{\bar z})}{(e_{\bar z},e_{\bar z})}=
\frac{(u_ne_{\bar z})(z)}{e_{\bar z}(z)}=\frac{u_n(1+z\bar z)^{n-1}}
{(1+z\bar z)^{n-1}}.\label{c}
\end{equation}

Observe that the symbol $s_n(u)$ is polynomial in $n$.

The adjoint action $Ad$ of $G$ on $su(2)$ (by rotations) can be naturally
ex\-tended to $U$.  Then for $u\in U$, $g\in G$ one has
$\pi_n(Ad(g)u)=\pi_n(g)\pi_n(u)\pi_n(g^{-1})$.
Therefore, from $G$-equivariance of covariant symbols it follows that
the mapping $u\mapsto s_n(u)$ is also $G$-equivariant, i.e., for $g\in G$
$s_n(Ad(g)u)(z,\bar z)=s_n(u)(zg,\overline{zg})$.

Now we will give an explicit description of the mapping $s_n$ using a
$G$-module structure of $U$ under adjoint action.

Consider elements of $sl(2,{\bf C})$ as complex linear functionals on
$su(2)$ with respect to $Ad$-invariant pairing $X,Y\mapsto -\frac{1}{2}trXY$
for $X\in sl(2,{\bf C})$ and $Y\in su(2)$. The symmetrization mapping $Sym$
(see \cite{D}) is a ${\bf C}$-linear isomorphism of the
algebra $\Lambda$ of all complex polynomials on $su(2)$ onto $U$ such that
if $f(Y)=-\frac{1}{2}trXY$ is a functional on $su(2)$ for an arbitrary $X\in sl
(2,{\bf C})$
then $Sym(f^k)=X^k$ for all natural $k$. It is $G$-equivariant, i.e., $Sym$
maps $f(Ad(g^{-1})Y)$ to $Ad(g)Sym(f)$ for $f\in\Lambda$. Let $I,M$ denote
the spaces of all rotation-invariant and harmonic polynomials in $\Lambda$
respectively.  Then $Z=Sym(I)$ is the center of $U$. Denote $E=Sym(M)$.
It is known that
$\Lambda=I\otimes M$ and $U=Z\otimes E$  (in the both tensor products
$x\otimes y$ corresponds to the respective product $xy$, see \cite{K}). Thus ea
ch
element $u\in U$ can be written as $u=z_1v_1+\dots+z_kv_k$ for some $z_i\in
Z, v_i\in E$.

Since $\pi_n$ is irreducible for each $z\in Z$ the operator $\pi_n(z)$
is scalar. Denote that scalar by $\chi_n(z)$. The function $\chi_n$
is a homomorphism of $Z$ into ${\bf C}$ and is called a central character
of $U$ corresponding to $\pi_n$.

\bigskip

Lemma 1. {\it For $z\in Z, u\in U$

$(i)$ the symbol $s_n(z)$ is a constant equal to $\chi_n(z)$;

$(ii)$ $s_n(zu)=s_n(z)s_n(u)$. }

\bigskip

Proof. Since the covariant symbol of the identity operator is identically 1,
the symbol of $\pi_n(z)$  is identically  equal to $\chi_n(z)$,
which proves $(i)$.  Now, $(ii)$ is obvious.

In order to describe $s_n$ on $U$ it suffices to know the restrictions
of $s_n$ to $Z$ and $E$.

The adjoint action of $G$ on $su(2)$ keeps invariant a square of
Euclidean radius, $(r(Y))^2=-\frac{1}{2}trY^2$, $Y\in su(2)$,
which is a quadratic polynomial on $su(2)$.
It is known (see, e.g.\cite{D}) that $Z$ is a polynomial
algebra in the Casimir element $z_0=Sym(r^2)$. A direct calculation
provides

\bigskip

Lemma 2. $s_n(z_0)=1-n^2$.

\bigskip

Let $M_k$ denote the subspace of $M$ of harmonic polynomials of degree $k$.
It is known that with respect to the action of $G$ on $\Lambda$,
via a change of variables, $M_k$ is a $(2k+1)$-dimensional irreducible
subspace.
Let $v_0$ denote the element of $U$ corresponding to
\begin{displaymath}
V=\left(\begin{array}{cc}
0&0\\1&0\end{array}\right)
\in sl(2,{\bf C}),
\end{displaymath}
$f_0(Y)=-\frac{1}{2}trVY$. Then for each natural $k$ $Sym(f_0^k)=v_0^k$.
It is easy to check directly that $f_0^k$ is harmonic, so $f_0^k\in M_k$.
Using (2) one gets that $(v_0^k)_n=(\frac{d}{dz})^k$ for all $n$.
Now from (3) follows

\bigskip

Lemma 3. $s_n(v_0^k)=(n-1)(n-2)\dots(n-k)(\frac{\bar z}{1+z\bar z})^k.$

\bigskip

Consider a $G$-equivariant embedding of $S^2$ in $su(2)$ given as follows
\begin{displaymath}
S^2\supset {\bf C}\ni z\mapsto\left(\begin{array}{cc}
i\frac{1-z\bar z}{1+z\bar z}&-2i\frac{\bar z}{1+z\bar z}\\
-2i\frac{z}{1+z\bar z}&-i\frac{1-z\bar z}{1+z\bar z}\end{array}\right)
\in su(2).
\end{displaymath}
The image of $S^2$ is the unit sphere with respect to the Euclidean scalar
product $X,Y\mapsto -\frac{1}{2}trXY$ in $su(2)$.
Then the pullback of $f_0(Y)$ to $S^2$ is
$i\frac{\bar z}{1+z\bar z}$. Thus identifying $S^2$ with the unit
sphere in $su(2)$ one gets
\begin{equation}
s_n(Sym(f_0^k))=s_n(v_0^k)=\frac{1}{i^k}(n-1)(n-2)\dots(n-k)f_0^k|_{S^2}.
\label{d}
\end{equation}

Since all the ingredients of (4) are $G$-equivariant, one can
replace $f_0^k$ in (4) by a linear combination of its rotations
by the elements of $G$.
Since $G$ acts irreducibly in $M_k$
one thus obtains an
arbitrary element of $M_k$.

\bigskip

Lemma 4. {\it For all $f\in M_k$}
$s_n(Sym(f))=\frac{1}{i^k}(n-1)(n-2)\dots(n-k)f|_{S^2}.$

\bigskip

Denote $E_k=Sym(M_k)$. Since $M=\oplus_kM_k$ then $E=\oplus_kE_k$.
Therefore, an arbitrary element of $U$ may be expressed as a sum
of monomials of a form $z_0^jv$ with $v\in E_k$. Combining Lemmas 1 - 4,
one gets

\bigskip
Proposition 1. {\it Let $v\in E_k$, $v=Sym(f)$ for some $f\in M_k$. Then}
$$
s_n(z_0^jv)=(\frac{1}{i})^{2j+k}(n^2-1)^j(n-1)(n-2)\dots(n-k)f|_{S^2}.
$$

\newpage

{\bf \S 3. Symbol algebras.}

\bigskip

Let $R$ denote the space of restrictions of all polynomials from $\Lambda$
to the unit sphere $S^2$. The elements of $R$ are called regular
functions on $S^2$. It is known (see, e.g.\cite{CH}) that
the restriction of the space $M$ of harmonic polynomials to $S^2$
is a bijection of $M$ onto $R$. Therefore each regular function on $S^2$
is a restriction of a unique harmonic polynomial.
Denote by $R_k$ the restriction of $M_k$.  Thus $R=\oplus_kR_k$.

Since for all $u\in U$ $s_n(u)$ is polynomial in $n$ one can consider
$s_t(u)$ for arbitrary $t\in {\bf C}$.
It is obvious that Lemma 2 is valid for  $s_t(u)$ for all complex $t$.
Namely the mapping $z\mapsto s_t(z)$ is a homomorphism of $Z$ to
${\bf C}$ and for $z\in Z,u\in U$ $s_t(zu)=s_t(z)s_t(u)$.

 Denote ${\cal A}_{1/t}=s_t(U)$. In the sequel ${\bf N}^\ast$ will
 denote the set
 of all positive integers.  From Proposition 1 immediately follows

\bigskip

Proposition 2. {\it For $t=n\in{\bf N}^\ast\ {\cal A}_{1/t}=
\oplus_{k=0}^{k=n-1}R_k$. For all other values of $\ t\quad
{\cal A}_{1/t}$ consists of all regular functions.}

\bigskip

We are going to show that the kernel of
the mapping $s_t$ is a
two-sided ideal in $U$, thus obtaining a quotient algebra structure
in ${\cal A}_{1/t}$.

Let $J_t$ be the two-sided ideal in $U$ generated by $Z\cap Ker\ s_t$.

\bigskip

Lemma 5. $U=E+J_t$.

\bigskip

Proof. For $u=zv$ with $z\in Z$, $v\in E$ one has
$u=s_t(z)v+(z-s_t(z))v$ where $s_t(z)$ is identified with
the respective constant in $U$. The assertion of Lemma follows from
the fact that $z-s_t(z)\in Z\cap  Ker\ s_t$.

\bigskip

Lemma 6. {\it For} $t\notin {\bf N}^\ast$ $ Ker\ s_t=J_t$.

\bigskip

Proof. From Lemma 5 follows that $Ker\ s_t=E\cap Ker \ s_t+J_t$.
Since the restriction of the space of harmonic polynomials to a sphere is
a bijection onto the space of regular functions,
 it follows from Lemma 4 that
$E\cap Ker\ s_t$ is trivial for $t\notin {\bf N}^\ast$.

\bigskip

Proposition 3. {\it For all $t\in {\bf C}$ $Ker\ s_t$ is a two-sided
ideal in $U$. }

\bigskip

Proof. For $t=n\in{\bf N}^\ast$ $Ker\ s_t=Ker\ \pi_n\subset U$.
For the rest of $t$ Lemma 6 is applied.

Now ${\cal A}_{1/t}$ carries a quotient algebra structure. Denote the
corresponding multiplication in ${\cal A}_{1/t}$ by $\ast_{1/t}$.

A following Lemma is obtained from direct calculations.

\bigskip

Lemma 7. {\it The function $f(z,\bar z)=(n+1)(n+2)\dots(n+k)
(\frac{\bar z}{1+z\bar z})^k$ is a contravariant symbol
of the operator $\pi_n(v_0^k)=(\frac{d}{dz})^k$ in $H_n$.}

\bigskip

Proposition 4. {\it For $n\in{\bf N}^\ast,\ u\in U$ the function
$s_{-n}(u)(-1/\bar z,-1/z)$ is a contravariant symbol of the operator
$\pi_n(u)$ in $H_n$.}

\bigskip

Proof. It follows from Lemma 2 that $s_n$ coincides with $s_{-n}$
on the center $Z$ of $U$. Therefore the ideals $J_n$ and $J_{-n}$
coincide as well. Since $J_n\subset Ker\ \pi_n$ for each $u\in J_{-n}$
both the symbol $s_{-n}(u)$ and operator $\pi_n(u)$ are zero.
It follows from Lemma 5, that it remains to check the assertion of the
Proposition for $u\in E_k$, since $E=\oplus E_k$.
The rest follows from Lemma 7, the irreducibility of $E_k$ with respect
to the adjoint action of $G$ and equivariance of contravariant symbols.

Thus the algebra ${\cal A}_{-1/n}$ consists of contravariant symbols
of all operators in $H_n$ up to the antipodal mapping $z\mapsto-1/\bar z$
of the sphere $S^2$. Moreover, the mapping which maps the symbol $s_{-n}(u)$
to the operator $\pi_n(u)$ in $H_n$ is a correctly defined
homomorphism of ${\cal A}_{-1/n}$ onto $End\ H_n$.

\bigskip

{\bf \S 4. The proof of the Correspondence Principle.}

\bigskip

Recall now some basic facts about filtration in
the universal enveloping algebra and Poisson
structure in the symmetric algebra of a Lie algebra (see, e.g.\cite{D}).

Let $U_k$ denote the subspace of $U$ spanned by monomials of
degree $\leq k$. Then $\{U_k\}$ is a filtration, for $u\in U_k$,
$v\in U_l$ both $uv,vu\in U_{k+l}$ and $uv-vu\in U_{k+l-1}$.

Let $\Lambda_k$ denote the subspace of $\Lambda$ of homogenous polynomials
of degree $k$. Then $Sym(\Lambda_k)\subset U_k$. Moreover, $Sym$ composed
with the quotient mapping $U_k\to U_k/U_{k-1}$ establishes an isomorphism
of $\Lambda_k$ onto $U_k/U_{k-1}$. For $u\in U_k$ let $\underline{u}$
denote the unique element of $\Lambda_k$ such that
$Sym(\underline{u})\equiv u\ {\rm mod}\ U_{k-1}$.

There exists a natural Poisson structure on $\Lambda$ such that
if $f_i(Y)=-\frac{1}{2}trX_iY,\ i=1,2,3$ are linear functionals on $su(2)$
corresponding to $X_i\in sl(2,{\bf C})$ with $[X_1,X_2]=X_3$,
then $\{f_1,f_2\}=f_3$. The symplectic leaves of that Poisson structure
are the $G$-orbits in $su(2)$, i.e. the spheres. Denote by
$\{\cdot,\cdot\}_{S^2}$ the restriction of the Poisson bracket to the unit
sphere $S^2$.  Then for $f,g\in\Lambda$
$\{f|_{S^2},g|_{S^2}\}_{S^2}=\{f,g\}|_{S^2}$.

For $u\in U_k,v\in U_l$ $\ \underline{uv}=\underline{vu}=
\underline{u}\cdot\underline{v}$ while
$\underline{uv-vu}=\{\underline{u},\underline{v}\}$.

\bigskip

Proposition 5. {\it Let $u\in U_k$. Then}
$$
\lim_{t\to\infty}\frac{1}{t^k}s_t(u)=\frac{1}{i^k}\underline{u}|_{S^2}.
$$

Proof. If $f\in\Lambda_k$ and $u=Sym(f)\in U_k$, then
$\underline{u}=f$. In particular $z_0\in U_2$, $\underline{z_0}=r^2$
and the restriction of $r^2$ to the unit sphere $S^2$ is
identically 1. Now the proof
follows from Proposition 1.

Let $f,g$ be regular functions on $S^2$. From Proposition 2
follows that for $t\notin{\bf N}^\ast$ or sufficiently big
$t=n\in {\bf N}^\ast$  the
product $f\ast_{1/t}g$ is defined.

\bigskip

Theorem. {\it For any regular functions $f,g$ on $S^2$ holds}
$$
 \lim_{t\to\infty}f\ast_{1/t}g=fg;
 \quad \lim_{t\to\infty}t(f\ast_{1/t}g-g\ast_{1/t}f)=i\{f,g\}_{S^2}.
$$
\bigskip

Proof. It is enough to consider $f\in R_k,\ g\in R_l$.
Let $u\in E_k,\ v\in E_l$ be such that $\frac{1}{i^k}\underline{u}$
and $\frac{1}{i^l}\underline{v}$ are the harmonic extensions of
$f$ and $g$, respectively. Then, using Proposition 5 one gets
$$
f\cdot g=\frac{1}{i^k}\ \underline{u}|_{S^2}\cdot
\frac{1}{i^l}\ \underline{v}|_{S^2}=\frac{1}{i^{k+l}}\ \underline{uv}|_{S^2}=
\lim_{t\to\infty}\frac{1}{t^{k+l}}s_t(uv)=
\lim_{t\to\infty}\frac{1}{t^{k+l}}\ s_t(u)\ast_{1/t}s_t(v).
$$
Applying Lemma 4 to the last expression one finally obtains
$$
f\cdot g=
\lim_{t\to\infty}\frac{(t-1)\dots(t-k)(t-1)\dots(t-l)}{t^{k+l}}f\ast_{1/t}g=
\lim_{t\to\infty}f\ast_{1/t}g.
$$
Proceeding in a similar manner one gets
$$
i\{f,g\}_{S^2}=i\{\frac{1}{i^k}\ \underline{u}|_{S^2},
\frac{1}{i^l}\ \underline{v}|_{S^2}\}_{S^2}=
\frac{1}{i^{k+l-1}}\{\underline{u},\underline{v}\}|_{S^2}=
\frac{1}{i^{k+l-1}}\ \underline{uv-vu}|_{S^2}=
$$
$$
\lim_{t\to\infty}\frac{1}{t^{k+l-1}}s_t(uv-vu)=
\lim_{t\to\infty}\frac{(t-1)\dots(t-k)(t-1)\dots(t-l)}{t^{k+l}}
t(f\ast_{1/t}g-g\ast_{1/t}f)=
$$
$$
\lim_{t\to\infty}t(f\ast_{1/t}g-g\ast_{1/t}f).\hskip 12 cm
$$

Let $F=\{1,1/2,1/3,\dots\}$. According to the Theorem,
for $h=1/n\in F$
the algebras ${\cal A}_{1/n}$ and ${\cal A}_{-1/n}$ of covariant and
contravariant symbols of operators in $H_n$ form the data for
covariant and contravariant quantization on $S^2$ respectively.

\bigskip

{\bf Acknowledgements}

\bigskip

I wish to express my gratitude to Professors R.G.Airapetyan,
B.V.Fedosov and M.S.Narasimhan for helpful discussions.
I am pleased to thank for kind hospitality the ICTP,
Trieste, where the work has been completed.
 
\end{document}